\documentclass[conference]{IEEEtran}
\IEEEoverridecommandlockouts
\usepackage{cite}
\usepackage{amsmath,amssymb,amsfonts}
\usepackage{graphicx}
\usepackage{textcomp}
\usepackage{xcolor}
\def\BibTeX{{\rm B\kern-.05em{\sc i\kern-.025em b}\kern-.08em
    T\kern-.1667em\lower.7ex\hbox{E}\kern-.125emX}}

\begin{document}

\title{RTL Interconnect Obfuscation By Polymorphic Switch Boxes For Secure Hardware Generation
}

\author{\IEEEauthorblockN{Haimanti Chakraborty and Ranga Vemuri}
\IEEEauthorblockA{\textit{Digital Design Environments Laboratory, Electrical and Computer Engineering Department} \\
\textit{University of Cincinnati}\\
Cincinnati, Ohio, USA \\
chakrahi@mail.uc.edu ; ranga.vemuri@uc.edu}

}

\maketitle

\begin{abstract}
Logic Obfuscation is a well renowned design-for-trust solution to protect an Integrated Circuit (IC) from unauthorized use and illegal overproduction by including key-gates to lock the design. This is particularly necessary for ICs manufactured at untrusted third-party foundries getting exposed to security threats. In the past, several logic obfuscation methodologies have been proposed that are vulnerable to attacks such as the Boolean Satisfiability Attack. Many of these techniques are implemented at the gate level that may involve expensive re-synthesis cycles. In this paper, we present an interconnect obfuscation scheme at the Register-Transfer Level (RTL) using Switch Boxes (SBs) constructed of Polymorphic Transistors. A polymorphic SB can be designed using the same transistor count as its Complementary-Metal-Oxide-Semiconductor based counterpart, thereby no increased area in comparison, but serving as an advantage in having more key-bit combinations for an attacker to correctly identify and unlock each polymorphic SB. Security-aware high-level synthesis algorithms have also been presented to increase RTL interconnects to Functional Units impacting multiple outputs such that when a polymorphic SB is strategically inserted, those outputs would be corrupted upon incorrect key-bit identification. Finally, we run the SMT (Satisfiability Modulo Theories)-based RTL Logic Attack on the obfuscated design to examine its robustness.
\end{abstract}


\section{Introduction}
Developing and maintaining billion-dollar fabrication facilities for the manufacture of Integrated Circuits (ICs) has become challenging with the scaling progression of the technology nodes \cite{pilato2018tao} \cite{citekey}. In order to save expenses, a number of semiconductor design houses are opting to become \textit{fab-less} and outsourcing their designs to third-party untrusted foundries that may potentially be located overseas \cite{pilato2018tao} \cite{citekey} \cite{chakraborty2023split} \cite{cui2019split}. Nonetheless, this has resulted in the increase in hardware security based threats \cite{chen2023guarding} \cite{chen2023work} by means of reverse engineering the functionality of an Intellectual Property (IP), illegal overproduction, piracy, counterfeiting of ICs by malevolent entities located at the untrusted foundry \cite{citekey} \cite{chen2023work} \cite{halder2023obnocs} \cite{halder2021low}. A study conducted by Business Action to Stop Counterfeiting and Piracy (BASCAP) has found the global value of counterfeit and pirated goods estimated to be \$600-650 billion \cite{verma2014economic} before 2015. 
\par In order to resolve this issue, several researchers in the recent years have proposed varied methodologies \cite{citekey} \cite{razaque2021analysis} to safeguard IC designs. Split Manufacturing, which was initially proposed for yield enhancement can also be utilized to protect an IC design \cite{vemuri2021split}. It essentially splits an IC into two parts: Front End Of Line (FEOL) comprising the lower metals and the transistors in the device layers, and Back End Of Line (BEOL) comprising the rest of the higher metal layers \cite{vemuri2021split} \cite{chakraborty2023split} \cite{citekey}. The FEOL portion can be manufactured at an untrusted foundry whereas the BEOL at a trusted facility, and then the two parts can be integrated together at the trusted facility \cite{vemuri2021split} \cite{chakraborty2023split} \cite{citekey}. The untrusted facility no longer having access to the full IC provides some design protection \cite{cui2019split} \cite{vemuri2021split} \cite{chakraborty2023split} \cite{citekey}. Watermarking is another strategy which is typically difficult to remove without destroying the IC functionality \cite{cui2019split}. Yet another technique is Logic Obfuscation (also known as Logic Encryption or Logic Locking) that obfuscates or hides a design's functionality by including some extra key gates to lock it \cite{citekey}. The design can only function correctly when the correct key bits (hidden from the untrusted foundry and stored in a tamper-proof secure memory) are identified \cite{pilato2018tao} \cite{citekey}. Numerous logic obfuscation methodologies have been proposed over the years at the gate-level abstraction level which may require expensive re-synthesis cycles \cite{cui2019split} \cite{vemuri2021split} \cite{citekey}. With enhanced complexity of an IC design, designers are migrating to higher abstraction levels for design automation \cite{pilato2018tao} that has relatively fewer design components \cite{vemuri2021split} \cite{citekey}. 
\par In this paper, we present an interconnect obfuscation methodology at a higher Register Transfer Level (RTL) of abstraction by incorporating Switch Boxes (SBs) designed using Polymorphic or Ambipolar Transistors. Polymorphic transistors exhibit the characteristics of ambipolarity \cite{citekey}, containing two gate terminals (Control Gate and the Polarity Gate) and a single device can either behave as an n-type or a p-type transistor depending on the voltage level supplied to the Polarity Gate \cite{de2012polarity}\cite{alasad2017logic} \cite{citekey}. Because of this feature, a polymorphic transistor can readily support camouflaging as well as logic encryption \cite{vemuri2021split} when the voltage values at the Control Gate and the Polarity Gate are hidden from the untrusted foundry as key-bits \cite{citekey}.  
\par We also present security-aware modified scheduling and allocation algorithms in high-level synthesis that would increase the Data Flow Graph edges to assign more RTL interconnects to specific RTL functional units that is able to impact multiple outputs. In doing so, when the polymorphic SBs are included in some of those locations subject to area overhead constraints, all of those outputs would be corrupted upon incorrect SB key-bit identification by an attacker to unlock the design. We finally run the SMT (Satisfiability Modulo Theories)-based RTL Logic Attack \cite{karfa2020register} on our obfuscated design to test for its resilience.

\section{Background}
In this section, we consider an example of a polymorphic device such as the Silicon Nanowire Field Effect Transistor (SiNW FET) \cite{de2012polarity} to discuss the phenomenon of ambipolarity \cite{citekey}. We then include logic obfuscation fundamentals and some prior work.
\subsection{The Phenomenon of Ambipolarity in SiNW FET Polymorphic Devices}
\begin{figure}[h]
\centering
\includegraphics[width=0.47\textwidth]{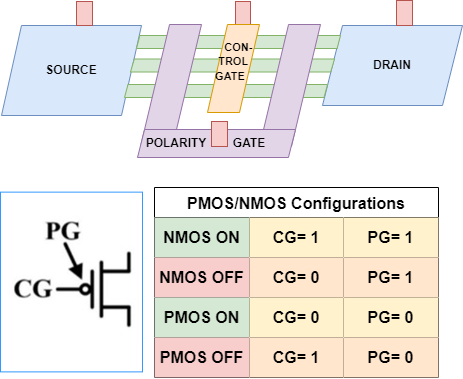}
\caption{Structural Representation of an Ambipolar SiNW FET that can be used as a PMOS or an NMOS Transistor based on the Control/Polarity Gate Logical Values \cite{de2012polarity} \cite{citekey}}
\label{figure1}
\end{figure}

The definition of \textit{Ambipolarity} may be expressed as the placement of both negative and positive charge carriers under bias constraints, that allows a designer to change device polarity \cite{alasad2017logic} \cite{citekey}. Based on the voltage value supplied to the device, its functionality can be changed from p-type to n-type and vice versa, and Schottky barriers are primarily responsible for it \cite{alasad2017logic} \cite{citekey}. The two-dimensional structure of a vertically-stacked gate-all-around (GAA) SiNW FET is demonstrated in Figure \ref{figure1}.

Figure \ref{figure1} demonstrates that a Silicon Nanowire FET device has two gates, namely, the Control Gate CG (that helps in turning the device off or on depending on the voltage value, similar to a traditional FET transistor) and the Polarity Gate PG (situated between the source and the drain, that helps modify the p-channel to the n-channel and vice versa) \cite{alasad2017logic} \cite{citekey}. When the voltage at CG is 1(0) and 0(1), the PMOS(NMOS) device is turned off and on respectively \cite{de2012polarity} \cite{citekey}. Additionally, if PG is high(low), the device behaves as an NMOS(PMOS) transistor respectively \cite{de2012polarity} \cite{citekey}. Although other emerging devices such as Carbon Nanotube (CNT) FETs, Nanoelectromechanical (NEM) relays and Graphene SymFETs have polarity control characteristics as well, SiNW FETs are compatible with the modern CMOS technology \cite{alasad2017logic} \cite{citekey}.
\subsection{Fundamentals of Logic Obfuscation/Encryption/Locking}
\begin{figure}
\centering
\includegraphics[width=0.5\textwidth]{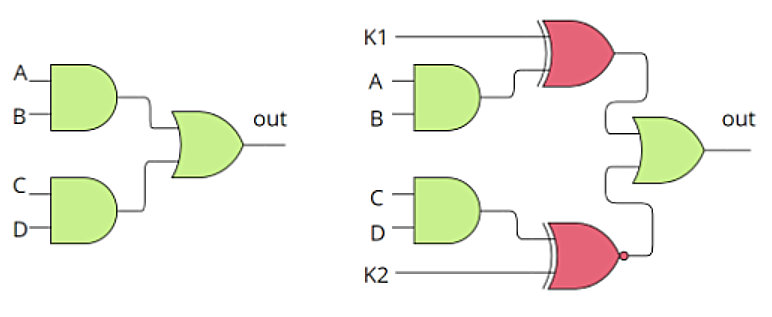}
\caption{Logic Encryption in CMOS Technology at the Gate-Level \cite{subramanyan2015evaluating} \cite{citekey}}
\label{figure2}
\end{figure}

Logic Obfuscation (or Logic Encryption or Logic Locking) is a potent hardware security methodology that can prohibit illegal overproduction and unauthorized use of an IC design \cite{citekey}. Figure \ref{figure2} showcases the primary principle behind logic obfuscation at the gate-level stage. The figure on the left is the actual design, whereas that on the right is the locked design with one XOR and one XNOR key gate (indicated in red color), and with K1 and K2 as the corresponding key bits \cite{citekey}. If the correct key-bits, in this case K1= 0 and K2= 1 are not applied, it would generate an incorrect functionality at the output \cite{citekey}. For real designs containing more key bits, the number of possible keys increase exponentially with the increase in the number of key gates, and unless the correct key is known, the IC cannot be used \cite{citekey}. This encrypted design is then sent to the untrusted foundry without supplying the correct key \cite{citekey}. The system integrator after receiving the fabricated IC, can place the correct key in a tamper-proof secure memory and integrate it with the IC for its correct functioning \cite{citekey}. 

\par
\begin{figure} [h]
\centering
\includegraphics[width=0.4\textwidth]{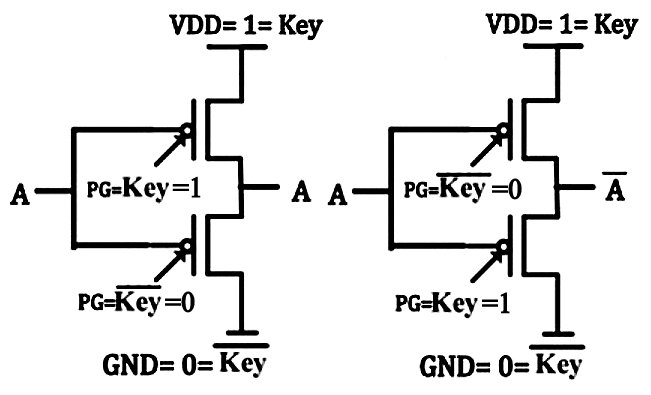}
\caption{Polymorphic Gate Based Logic Obfuscation with the same circuit behaving as a Buffer on the left and an Inverter on the right with change in Polarity Gate (PG) values \cite{de2012polarity} \cite{alasad2017logic}\cite{citekey}}
\label{figure3}
\end{figure}

 Figure \ref{figure3} demonstrates how one is able to perform logic obfuscation and camouflaging with the help of polymorphic transistors. Keeping the standard values of GND and VDD as 0 and 1 respectively, if the PG values of the lower and the upper FETs are 0 and 1 respectively, the two-transistor circuit behaves as a polymorphic buffer \cite{de2012polarity} \cite{citekey}. Whereas if we reverse the polarities of the two PG values, or if the supply values are reversed, the same circuit behaves as an inverter \cite{de2012polarity} \cite{citekey}. Hence, if these PG values and/or the values at the supply rails are hidden as key bits, the desired functionality will not be obtained if incorrect key values are supplied \cite{citekey}. We will utilize this feature of ambipolar, polymorphic devices to our advantage to build polymorphic switch boxes (SBs) for the RTL Interconnect Obfuscation, subject to area constraints allowed by the design. We would aim to hide the CG and the PG values of each polymorphic transistor in the SB as key-bits, and if correct values are not applied, it would corrupt the correct connectivity of the interconnects, thereby corrupting the output functionality as well.

\subsection{Prior Work}
Although several logic encryption techniques have been implemented at the gate-level abstraction level, some prior research has been conducted at higher abstraction levels as well \cite{citekey}. Pilato et al. proposed the \textit{Techniques for Algorithm-Level Obfuscation, or TAO} obfuscation method in high-level synthesis \cite{pilato2018tao} to lock the control flow branches, constants, operators and dependencies available in that abstraction level. 
Rajendran et al. proposed the \textit{Belling the CAD} method \cite{rajendran2015belling} that added decoy connections into an ESL (Electronic System Level) tool-generated design to counter reverse engineering. Cui et al. proposed an RTL Obfuscation Method based on Split Manufacturing \cite{cui2019split} that included dummy multiplexers (MUXes) and Functional Units (FUs) for obfuscation besides delegating some interconnections to the BEOL layers. Wu et al. \cite{wu2015tpad} utilized a standard CMOS (Complementary Metal Oxide Semiconductor) based switch box for logic obfuscation at the gate level design stage. Such a switch box has fewer pass transistor key-bits to help identify if the correct connection would be of type crisscross "cross" or the parallel "bar", which may be relatively easier for an attacker to identify. Additionally, a gate-level based obfuscation may incur expensive re-synthesis cycles \cite{citekey} as compared to higher abstraction level based obfuscations that contain fewer hardware components.

\section{Attack Model}
The Threat Model used in this paper is the SMT (Satisfiability Modulo Theories)-based RTL Logic Attack as discussed in \cite{karfa2020register}. In order for the attack algorithm to determine the secret key of a locked RTL design, it models the RTL design as an RTL finite state machine with datapath (RTL-FSMD) \cite{karfa2020register} \cite{citekey} by applying the rewriting approach in \cite{karfa2010verification}. The authors \cite{karfa2020register} abstract out the details of the hardware into a behavioral program on which they launch the attack \cite{citekey}. This attack finds Distinguishing Input Patterns (DIPs) iteratively, similar to the Satisfiability (SAT) Attack \cite{subramanyan2015evaluating}, to eliminate equivalence classes of incorrect keys, and terminates when no DIPs are found \cite{karfa2020register} \cite{citekey}. For details, we refer the readers to \cite{karfa2020register}.

\section{RTL Interconnect Obfuscation Methodology}
In this section, we discuss a simple polymorphic switch box (SB) design using polymorphic transistors for RTL Interconnect Obfuscation. We also include modified high-level synthesis algorithms for scheduling and allocation of Data Flow Graph nodes and edges to assign them to RTL functional units which would impact multiple outputs such that when the polymorphic SBs are included at those specific locations, they would corrupt all of those outputs when an attacker incorrectly identifies the polymorphic SB key-bits to unlock them. Finally, we test the obfuscated design against the SMT-based RTL Logic Attack \cite{karfa2020register} to evaluate the robustness of the obfuscation method.

\subsection{Polymorphic Switch Boxes from Polymorphic Transistors}
\begin{figure} [h]
\centering
\includegraphics[width=0.4\textwidth]{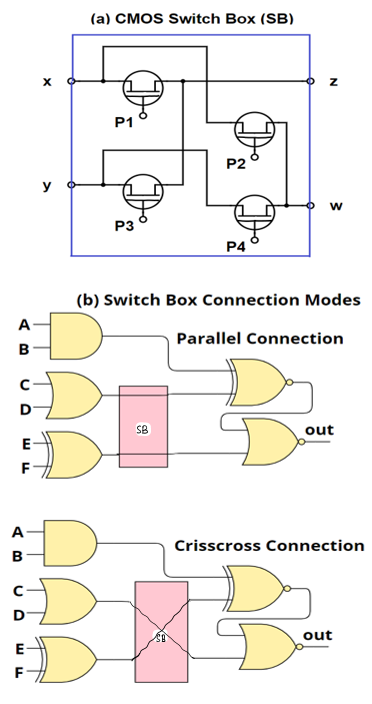}
\caption{(a) Standard Switch Box using 4 NMOS Pass Transistors; (b) Parallel and Crisscross Connection Modes using a Switch Box at Gate Level \cite{wu2015tpad}}
\label{figure4}
\end{figure}

Polymorphic Electronics was established by A. Stoica's group \cite{stoica2001polymorphic} at the NASA Jet Propulsion Laboratory as a novel class of electronic devices for a new approach to reconfiguration \cite{gajda2011evolutionary}. 
Switch boxes using standard CMOS (Complementary Metal Oxide Semiconductor) transistors for obfuscation has been proposed in the past \cite{wu2015tpad}. Figure \ref{figure4}(a) shows a CMOS based switch box (SB) composed of 4 NMOS pass transistors (and SRAMs to store the programming bits) and Figure \ref{figure4}(b) shows the two modes (crisscross and parallel) in which it can be configured to obfuscate a gate-level design as implemented in \cite{wu2015tpad}. In Figure \ref{figure4}(a), if P1P2P3P4 key vector= 1001, it would be a parallel SB connection (that is, with x connected to z and y connected to w) and if P1P2P3P4 key vector= 0110, we would get a crisscross SB connection (that is, with x connected to w and y connected to z). Out of the two modes, only one would be correct and any other bit combinations would shuffle the correct interconnect connectivity. Since this configuration has 4 key bits to identify a connectivity, the total number of possible key bit combinations for an attacker to identify the correct key vector is $2^{4}$ = 16, out of which one key vector is responsible for the parallel connection and another for the crisscross connection.  

\begin{figure} [h]
\centering
\includegraphics[width=0.38\textwidth]{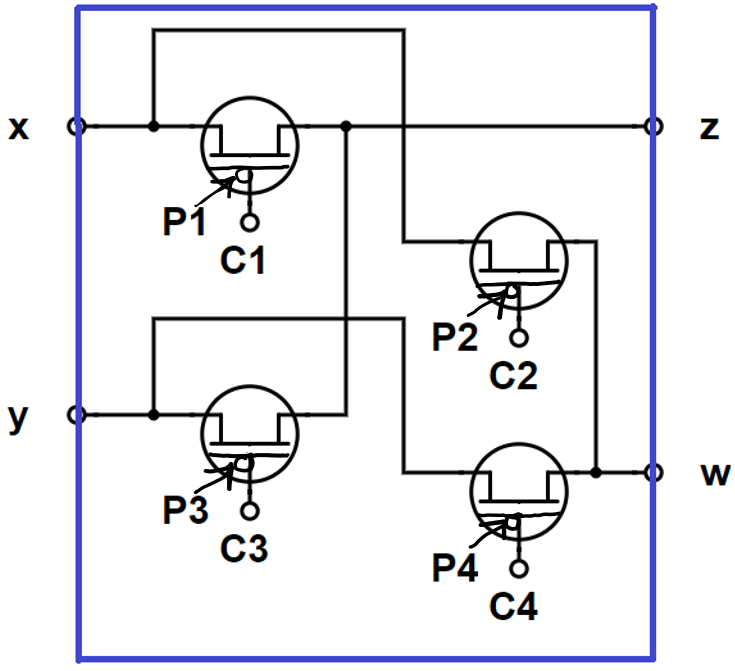}
\caption{The 4 Polymorphic Transistor Based Switch Box}
\label{figure5}
\end{figure}
\par Now if we replace all the 4 CMOS pass transistors with 4 polymorphic transistors, each transistor would now have a Control Gate (CG) key bit value and a Polarity Gate (PG) key bit value, thus doubling the number of key bits per transistor, resulting in more effort required by an attacker to guess the correct key bits per polymorphic SB. Figure \ref{figure5} showcases a Polymorphic SB configuration containing 4 Polymorphic transistors, where P1, P2, P3, P4 are the Polarity Gates of the four polymorphic transistors, and C1, C2, C3, C4 are the four Control Gates. From Figure \ref{figure1}, we know the CG and the PG values for when the polymorphic transistor would behave as a PMOS (off/on) or an NMOS (off/on) transistor. Considering these values as key bits, if in Figure \ref{figure5}, we have C1P1C2P2= 0010 or 1101 or 0001 or 1110, X would be connected to Z, and if C3P3C4P4= 1000 or 0111 or 1011 or 0100, Y would be connected to W, making it a parallel connection. Next, if we have C1P1C2P2= 1000 or 0111 or 1011 or 0100, X would be connected to W, and when C3P3C4P4= 0010 or 1101 or 0001 or 1110, Y would be connected to Z, making it a crisscross connection. Although it appears that we can have more than one key vector to make a parallel or a crisscross connection, the total number of key bit combinations per polymorphic SB has now increased from 16 (CMOS SB) to $2^{8}$ = 256 possible combinations without increase in the area cost from that of the CMOS SB counterpart. Any other key bit combinations would shuffle the correct interconnect connectivity. If we insert more than one polymorphic SB (say, x) subject to allowable area constraints, the attack complexity (interconnect wire combinations) would increase exponentially to $2^{x}$ \cite{wu2015tpad} with each polymorphic SB having 256 possible key bit combinations.
\par Section IV-C discusses that we can configure some SBs as parallel and some SBs as crisscross to place them in strategic locations to confuse an attacker. By strategic locations, one way to achieve it would be to insert a polymorphic SB between two interconnects which enter an RTL functional unit that fans out its output results to more than one register. This has been shown in Figure \ref{figure6}. That way, upon incorrect identification of the polymorphic SB key-bits, all such output register values would be corrupted. We can attain this by assigning specific nodes and edges of the design at the Data Flow Graph stage to RTL functional units via security-aware scheduling and allocation (high level synthesis) to generate the RTL. The following section IV-B discusses this.

\subsection{High-Level Synthesis (HLS) and Security-Aware HLS Algorithms}
High Level Synthesis (HLS) generates a design specified in a high level language (such as a Hardware Description Language or C/C++, etc.) into its equivalent hardware at the Register-Transfer Level (RTL) containing registers, multiplexers, functional units (FUs) such as adders, subtractors, multipliers, etc. \cite{vemuri1993experiences} \cite{ray2023redacting} \cite{citekey} \cite{chakraborty2023split}. HLS involves several steps including Data Flow Graph (DFG) construction, then operates on the DFG to perform scheduling, allocation, binding, and datapath and controller synthesis to generate the final structured RTL design \cite{vemuri1993experiences} \cite{citekey} \cite{chakraborty2023split}. The RTL design then goes through logical and physical synthesis to generate the design layout for fabrication \cite{citekey} \cite{chakraborty2023split}.
\par
\begin{figure} [h]
\centering
\includegraphics[width=0.49\textwidth]{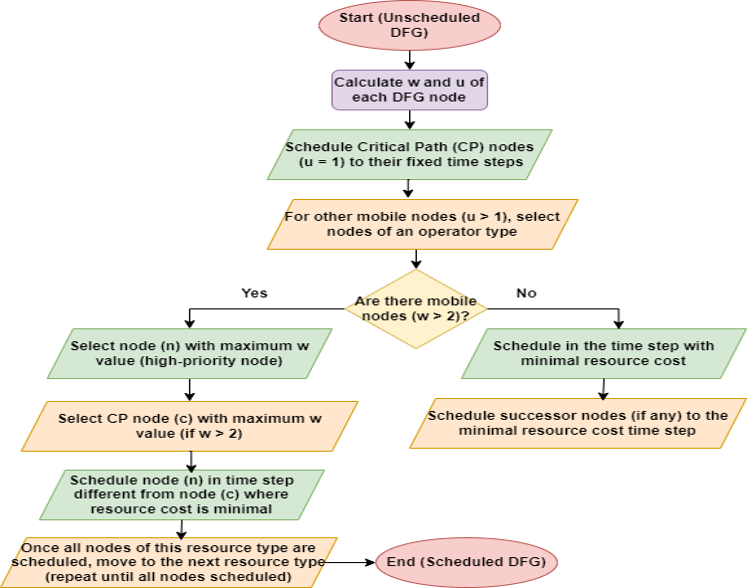}
\caption{Security-Aware Scheduling Flowchart \cite{paulin1989force} \cite{citekey}}
\label{figure10}
\end{figure}

\begin{figure} [h]
\centering
\includegraphics[width=0.4\textwidth]{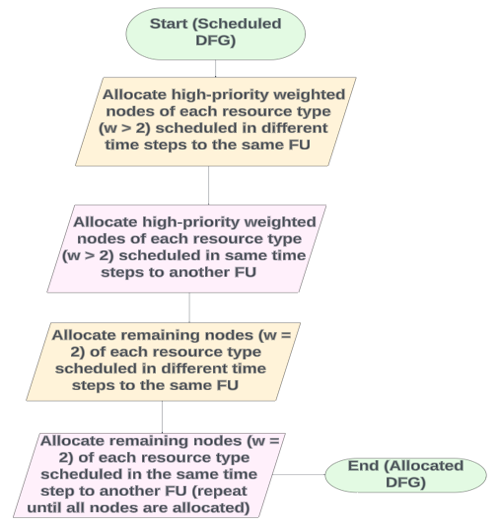}
\caption{Security-Aware Allocation Flowchart \cite{paulin1989force} \cite{citekey}}
\label{figure11}
\end{figure}

\begin{figure} [h]
\centering
\includegraphics[width=0.49\textwidth]{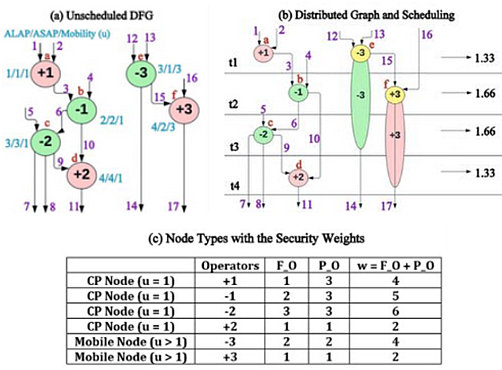}
\caption{Modified Scheduling and Allocation \cite{paulin1989force} \cite{citekey}}
\label{figure8}
\end{figure}

 Algorithms 1 and 2 (represented as flowcharts in Figures \ref{figure10} and \ref{figure11} respectively)  and Figure \ref{figure8} show the modified security-aware scheduling and allocation algorithms employed in this methodology. Chakraborty et al. proposed security-aware scheduling and allocation algorithms for RTL based obfuscation in \cite{citekey}. However, those are suitable for two operators/resource types at a time. In this paper, we present modified HLS algorithms for security that would consider one resource type at a time.

Algorithm 1 in Figure \ref{figure10} begins by computing the security-weight \textit{w} and scheduling mobility $\mu$ of each DFG node \cite{citekey}. \textit{w} is an arithmetic sum of two metrics $P_O$ and $F_O$, where $P_O$ is the number of primary outputs a particular node has paths ending in, and $F_O$ is the fanout count or the number of output dependencies/edges of that node to the next node(s) \cite{citekey}. The more the values of $P_O$ and/or $F_O$, the more the number of register values would be corrupted from the outputs of a functional unit at the RTL stage \cite{citekey} upon incorrect identification of the polymorphic SB key-bits inserted between two interconnects entering that functional unit. $\mu$ indicates the scheduling range of a node, subject to meeting design timing constraints, and is calculated using the formula: [As-Late-As-Possible (ALAP) schedule time of the node - As-Soon-As-Possible (ASAP) schedule time of the node + 1] \cite{paulin1989force} \cite{citekey}. The next step is to schedule all the nodes belonging to the critical path (CP) ($\mu$ = 1, hence these nodes have a fixed time step for scheduling) to meet design timing \cite{citekey}. The remainder of the algorithm deals with scheduling all remaining nodes with $\mu$ $>$ 1 \cite{citekey}. Selecting one operator type at a time from this list, if there are nodes with \textit{w} $>$ 2, that is, either one of $P_O$, $F_O$ or both are $>$ 1 (and hence will corrupt multiple RTL registers upon incorrect polymorphic SB key identification), we would give them higher priority \cite{citekey}. Next, if we have CP nodes with \textit{w} $>$ 2 as well, we select the CP node with the maximum \textit{w} value \cite{citekey}. And we schedule the high-priority node to the time step different from this CP node, where resource cost remains minimal (as discussed in \cite{paulin1989force}). Being in different time steps, both these weighted nodes (affecting multiple RTL registers) could be assigned the same FU for greater corruptibility \cite{citekey}. If there are nodes with \textit{w} = 2, that is, both $P_O$ and $F_O$ are 1 each, they would be scheduled in the time step corresponding to the least resource cost \cite{paulin1989force} \cite{citekey}. Once scheduled, we schedule the successor node(s) (if any) of the most recently scheduled node, and place it in the time step with minimal resource cost \cite{paulin1989force} \cite{citekey}. Once the operators of the first resource type are scheduled, we move to the next resource type and continue until nodes of all the resource types are scheduled \cite{paulin1989force} \cite{citekey}. 

\par Algorithm 2 in Figure \ref{figure11} assigns all high-priority weighted nodes \textit{w} $\geq$ 2 of the same resource type and scheduled in different time steps to the same FU \cite{citekey}. Other nodes of the same resource type belonging to overlapping time steps would be assigned a different FU of the same resource type \cite{citekey}. This is continued until all resources are allocated to the FUs \cite{citekey}.

\par Figure \ref{figure8} from \cite{citekey} demonstrates the aforementioned scheduling and allocation methods. Figure (a) shows the unscheduled DFG with ALAP/ASAP/$\mu$ values in blue for each node, node names from a-f and edge numbers 1-17 \cite{citekey}. Figure (b) shows the Distribution Graph of each node from time steps t1-t4 with the initial resource cost indicated on the right for each step \cite{paulin1989force} \cite{citekey}. Since $\mu$ of nodes e and f are 3 each, they are shown across three time steps \cite{citekey}. Figure (c) lists the security weights based on $P_O$ and $F_O$ \cite{citekey}. 

Out of the two mobile nodes, '-3' weighs the most (=4), so we select that as a high-priority node \cite{citekey}. In the CP node list, '-2' has the maximum weight (=6), so we pick that as well \cite{citekey}. Since CP nodes have a fixed time step ('-2' scheduled to time step t3), we would like '-3' to be scheduled to a different step from that (in order to allocate to the same FU) \cite{citekey}. So we eliminate t3 from '-3', which leaves us with t1 and t2 as the possible scheduling steps for '-3' \cite{citekey}. To find the minimal cost location, we use the Force-Directed Scheduling formula in \cite{paulin1989force} and find t1 to be a good location for '-3' \cite{citekey}. Now, '+3' is a successor of '-3' (because of the dependency relation) \cite{paulin1989force} \cite{citekey}. We eliminate t4 as a choice since '+2' CP node is scheduled over there and we are looking to schedule same resource type in different time steps to be allocated the same FU. Which then leaves us with t2 and t3 as choices for '+3' (t1 is not an option for '+3' since its predecessor node '-3' is already scheduled there) \cite{citekey}. From the same formula in \cite{paulin1989force} again, we find that the cost minimal location for '+3' is same for both t2 and t3, so we pick t2 (either one of t2 or t3 could have been picked in this case) \cite{citekey}. The yellow colored circles for '-3' and '+3' in Figure 6(b) indicate the final scheduling locations for these two nodes \cite{citekey}. We repeat the steps until all nodes are scheduled for each resource type.
\par For Allocation, we know from prior knowledge that nodes scheduled in different time steps can be allocated to the same FU \cite{citekey}. In this example, we allocate the nodes '-1', '-2', '-3' to the same subtractor-type FU, and '+1', '+2', '+3' to the same adder-type FU. In this way, the process is repeated until all nodes are allocated in a given design.

\subsection{Strategic Locations for The Polymorphic Switch Box Insertion for RTL Interconnect Obfuscation}
The previous Section IV-B helped to increase high priority weighted nodes to be assigned the same FU provided they could be scheduled in different time steps subject to design latency bound/timing constraints. The weights were computed based on nodes having paths ending in more than one primary output and/or if they had fanout of more than one to other nodes \cite{citekey}. From prior knowledge, we know that if DFG edges cross atleast one common time step boundary (that is, if they have overlapping time step boundary), they need to be assigned different registers. Hence, when such high priority nodes are maximized to be allocated to the same RTL FU subject to design timing constraints, the combined weight factor ensures that more output registers can be corrupted if an obfuscator (polymorphic SB, in this case) is placed between two interconnects that enter such an FU, and incorrectly unlocked with wrong keys. We consider such a location as one of the strategic ones for the polymorphic SB insertion.

\begin{figure} [t]
\centering
\includegraphics[width=0.49\textwidth]{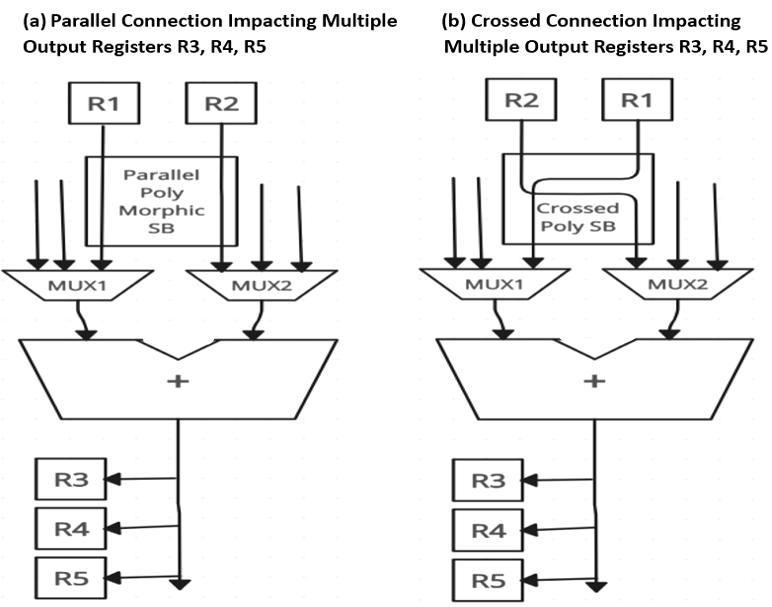}
\caption{Strategic Polymorphic Switch Box Insertion for RTL Interconnect Obfuscation, Parallel and Crossed Connections Impacting Multiple Output Registers R3, R4, R5}
\label{figure6}
\end{figure}

\par Another strategic location is when a polymorphic FU is inserted between two interconnects entering two FUs of different resource types (say, one being an adder FU and another one being a multiplier FU, etc.). That way, if the polymorphic SBs are again incorrectly unlocked with wrong keys, the functionality of the intended design would change.
\par
\begin{figure} [t]
\centering
\includegraphics[width=0.5\textwidth]{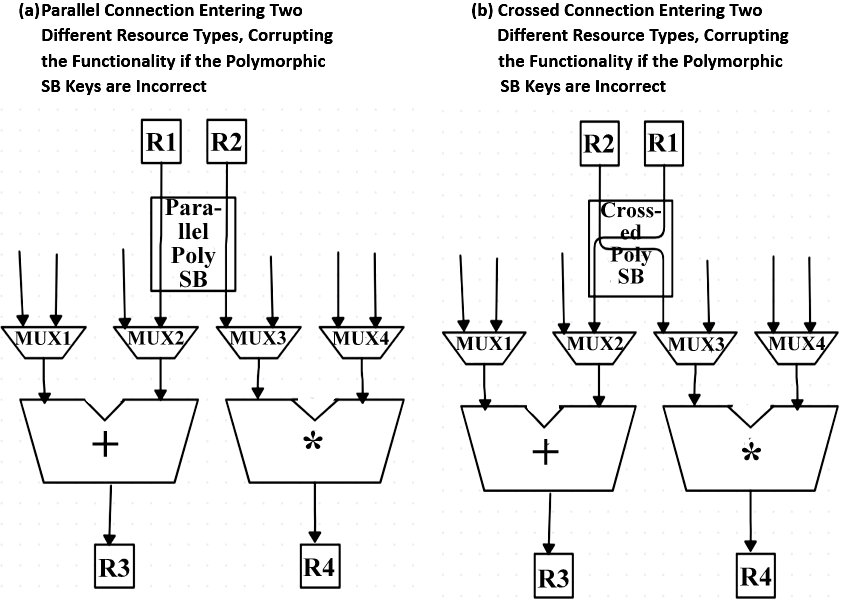}
\caption{Strategic Polymorphic Switch Box Insertion for RTL Interconnect Obfuscation, Parallel and Crossed Connections Entering Two Different Resource Types and Corrupting the Functionality if the Polymorphic SB Keys are Incorrect}
\label{figure6a}
\end{figure}

 Subject to allowable area constraints, the more polymorphic SBs we are able to include, the output corruptibility would increase. We propose using a mix of a few parallel type polymorphic SBs and a few of the crisscross variant to increase confusion for an attacker. These are illustrated in Figures \ref{figure6} and \ref{figure6a}.

\section{Experimentation and Results}
\begin{figure} [h]
\centering
\includegraphics[width=0.5\textwidth]{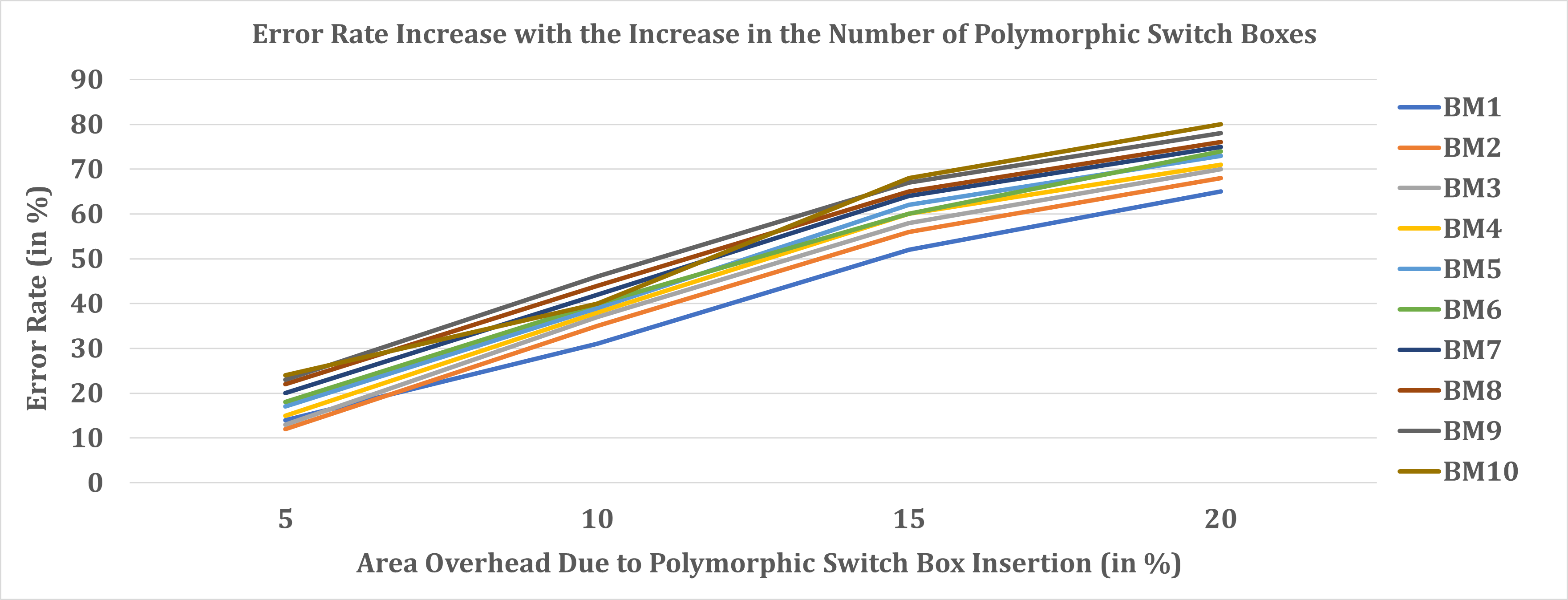}
\caption{Error Rate Versus Percentage of Area Overhead with Polymorphic Switch Box Insertion}
\label{figure7}
\end{figure}

We evaluate our method using benchmarks from \cite{WinNT} and internet sources on a system with AMD Ryzen 7 2700X Eight-Core Processor and 15GB memory. \newline \indent
Once the security-aware scheduling and allocation are performed for all the operators based on security-aware weights, the RTL datapaths and controllers are obtained for each benchmark circuit \cite{citekey}. Then depending on the allowed area constraints, polymorphic SBs of the parallel and crisscross variants are inserted at the strategic locations discussed in Section IV-C. We then run the SMT based attack \cite{karfa2020register} on the RTL interconnect obfuscated designs.
We measure for different benchmarks the output error rate versus the percentage of area overhead due to addition of the polymorphic SBs. The results are shown in Figure \ref{figure7}. It can be seen that the error rate is increasing with the increase in the number of polymorphic SBs (represented as the percentage of area overhead increase).

\begin{table} [h!]
\caption{Benchmark Details: Schedule Length, Operator Count and Edge Count}
\begin{center}
\begin{tabular}{c c c c }
Benchmark & Schedule Length & Operator Count & Edge Count \\
 \hline
BM1 & 40 & 202 & 405\\
\hline
BM2 & 45 & 250 & 451\\
\hline
BM3 & 55 & 261 & 457\\
\hline
BM4 & 60 & 265 & 478\\
\hline
BM5 & 68 & 271 & 482\\
\hline
BM6 & 79 & 275 & 489\\
\hline
BM7 & 85 & 301 & 501\\
\hline
BM8 & 101 & 354 & 505\\
\hline
BM9 & 110 & 408 & 510\\
\hline
BM10 & 112 & 507 & 521\\
\hline

  \end{tabular}
  \label{tab1}
  \end{center}
  \end{table}
  
\begin{table} [h!]
\caption{Benchmark Details: Primary Output Count, Number of Polymorphic Switch Boxes Used and Total Key Bit Size}
\begin{center}
\begin{tabular}{c c c c}
Benchmark & Primary Outputs & Polymorphic SB Count & Key Size \\
 \hline
BM1 & 10 & 16 & 128 bits \\
\hline
BM2 & 17 & 20 & 160 bits\\
\hline
BM3 & 25 & 23 & 184 bits\\
\hline
BM4 & 36 & 25 & 200 bits\\
\hline
BM5 & 41 & 28 & 224 bits\\
\hline
BM6 & 50 & 31 & 248 bits\\
\hline
BM7 & 55 & 35 & 280 bits\\
\hline
BM8 & 67 & 39 & 312 bits\\
\hline
BM9 & 78 & 55 & 440 bits\\
\hline
BM10 & 86 & 64 & 512 bits\\
\hline

  \end{tabular}
  \label{tab2}
  \end{center}
  \end{table}

Tables \ref{tab1} and \ref{tab2} show the details pertaining to different benchmarks we used during the experiment. We tested the resilience of the RTL interconnect obfuscation method against the SMT based attack \cite{karfa2020register} and found that for each of the ten benchmarks obfuscated with the 20 percent area overhead and using a mix of the 4 encryption configurations as shown in Figures \ref{figure6} and \ref{figure6a}, it timed out after 10 hours without deciphering the key.

\section{Conclusion}
In this paper, we presented an RTL Interconnect Obfuscation method by using polymorphic transistors to construct Polymorphic Switch Boxes (SBs). We also presented modified security-aware scheduling and allocation algorithms at high-level synthesis to increase the RTL interconnect count to RTL functional units that would impact multiple outputs, such that when polymorphic SBs are inserted in strategic locations subject to area constraints, they would corrupt those outputs upon incorrect key-bit identification to unlock the SBs. We finally tested the robustness of the obfuscation method against the SMT-based RTL Logic Attack and found the method to be resilient.

\bibliographystyle{ieeetr}
\bibliography{main}

\end{document}